\documentclass[sigconf]{acmart}

\renewcommand\footnotetextcopyrightpermission[1]{}

\AtBeginDocument{%
  }

  \newcommand{\inlinequote}[1]{
{\textit{``#1''}}}
\newif\ifdraft
\draftfalse 

\usepackage{xcolor}

\newcommand{\MyBox}[1]{%
  \vspace{3mm}
  \noindent
  \colorbox{blue!5}{
    \parbox{0.97\linewidth}{#1}%
  }%
  \vspace{2.5mm}
}

\renewcommand\footnotetextcopyrightpermission[1]{} 

\begin{document}
\makeatletter
\fancyhead{} 
\makeatother

\title{Why Do We Code? A Theory on Motivations and Challenges in Software Engineering from Education to Practice}

\author{Aaliyah Chang}
\email{aaliyah.chang@queensu.ca}
\orcid{0009-0003-0891-013X}
\affiliation{
  \institution{Queen's University}
  \city{Kingston}
  \state{Ontario}
  \country{Canada}
  }

\author{Mariam Guizani}
\email{mariam.guizani@queensu.ca}
\orcid{0000-0003-2545-2612}
\affiliation{
  \institution{Queen's University}
  \city{Kingston}
  \state{Ontario}
  \country{Canada}
  }

\author{Brittany Johnson}
\email{johnsonb@gmu.edu}
\orcid{0000-0002-0271-9647}
\affiliation{
  \institution{George Mason University}
  \city{Fairfax}
  \state{Virginia}
  \country{USA}
  }

\renewcommand{\shortauthors}{Chang, Guizani and Johnson}

\begin{abstract}
Motivations and challenges jointly shape how individuals enter, persist, and evolve within software engineering (SE), yet their interplay remains underexplored across the transition from education to professional practice.
We conducted 15 semi-structured interviews and employed the Gioia Methodology, an adapted grounded theory methodology from organizational behaviour, to inductively derive taxonomies of motivations and challenges, and build the \textit{Exposure–Pursuit–Evaluation (EPE)} Process Model. Our findings reveal that impactful early exposure triggers intrinsic motivations, while non-impactful exposure requires an extrinsic push (e.g., career/ personal goals, external validation). We identify curiosity and avoiding alternatives as a distinct educational drivers, and barriers to belonging as the only challenge persisting across education and career. Our findings show that career progression challenges (e.g., navigating the corporate world) constrain extrinsic fulfillment while technical training challenges, barriers to belonging and threats to motivation constrain intrinsic fulfillment. The theory shows how unmet motivations and recurring challenges influence persistence, career shifts, or departure from the field. Our results provide a grounded model for designing interventions that strengthen intrinsic fulfillment and reduce systemic barriers in SE education and practice.

\end{abstract}

\maketitle
\section{Introduction}

Understanding why individuals choose to pursue and remain in software engineering (SE) is critical for informing education, workforce development, and retention strategies. Motivations, challenges, and their interplay are central to how people navigate their educational and professional pathways in SE. Yet, despite growing attention to affect \cite{graziotin2018happens, serebrenik2017emotional, girardi2021emotions, kurian2023importance, graziotin2019happiness}, well-being \cite{godliauskas2025well, russo2024understanding, aires2024exploring, takaoka2024exploring, wong2023mental, de2024hybrid}, and work satisfaction \cite{aires2024exploring, johnson2019effect, stol2022gamification, hussain2024exploring}, we still know little about how these two dimensions jointly shape experiences across the full SE journey, from education to professional practice.

A seminal 2008 systematic literature review \cite{beecham2008motivation} established a foundational understanding of motivations in SE. However, it predates several shifts in the field, including hybrid work arrangements, diversity, equity, and inclusion initiatives, and the introduction of artificial intelligence into development practice. These changes underscore the need to re-examine contemporary motivations. More recent studies examine career-specific motivations within particular organizational contexts, such as a not-for-profit R\&D organization \cite{FrancaMotivation2018} and other Brazilian organizations \cite{FrancaMotivation2020}. Challenges faced by software engineers have also been investigated through focused lenses, including burnout \cite{tulili2023burnout, trinkenreich2023model, paula2024burnout}, imposter phenomenon \cite{chen2024impostor, rosenstein2020identifying, guenes2024impostor}, and unhappiness \cite{graziotin2017unhappiness, graziotin2017consequences, obi2025identifying}.

However, existing work examines motivations and challenges separately and across distinct stages of the journey (education or career). This leaves an important gap in understanding how these dimensions co-occur, interact, and shape persistence and adaptation across the education–career continuum. To address this, we ask:

\begin{itemize}
\item \textbf{RQ1:} What are the motivations to pursue an education and career in software engineering?
\item \textbf{RQ2:} What are the challenges experienced by software engineers in their education and careers?
\item \textbf{RQ3:} How do motivations and challenges interact throughout the software engineering journey?
\end{itemize}

We conducted semi-structured interviews with 15 participants who had experienced both SE education and professional practice, spanning diverse roles and backgrounds. Using the Gioia Methodology \cite{gioia2013seeking}, we inductively derived first-order concepts, second-order themes, and aggregate dimensions to build a theory of motivation–challenge interplay in SE from education to practice.

Our work contributes the following: 
\begin{enumerate}
\item A comprehensive, empirically grounded taxonomy of motivations that lead individuals to pursue and remain in SE, spanning educational and professional contexts.
\item A taxonomy of challenges faced by software engineers across their education and career contexts, including barriers to belonging, career-progression obstacles, and threats to motivation.
\item A process theory explaining how motivations and challenges interact to shape individual trajectories within SE.
\end{enumerate}

By theorizing the motivation–challenge interplay across education and career, this work extends existing understandings of developer experience and provides insights for educators, organizations, and researchers seeking to attract, support, and retain a resilient SE workforce.

\section{Research Method}
\label{sec:method}
\begin{table*}
\caption{Interview participants' background information}
\footnotesize
\label{table:participants}
\begin{tabular}{lllllll}
\toprule
{\textbf{ID}} & {\textbf{Role}} & {\textbf{Gender}} & {\textbf{Age}} & \textbf{Country of Residence}  &  \textbf{Years of Experience} & {\textbf{Highest Level of Education}}\\\hline
P1 & Junior Software Developer & Man & 18-24 Years & Canada & 3-5 Years& Undergraduate Degree\\
P2 & Embedded Software Engineer & Woman & 18-24 Years & Canada & Less than 1 Year & Undergraduate Degree\\
P3 & Full Stack Developer & Woman & 18-24 Years  & Canada & Less than 1 Year & Undergraduate Degree\\
P4 & Principal Technical Program Manager & Woman & 55-64 Years & Canada & Over 10 Years & Technical Training\\
P5 & Senior Full Stack Software Engineer & Man & 25-34 Years & United States of America & 6-10 Years & Undergraduate Degree\\
P6 & CTO & Man & 25-34 Years & Germany & Over 10 Years & Master's Degree\\
P7 & Technical Architect & Woman & 25-34 Years  & United States of America & 3-5 Years & Master's Degree\\
P8 & Junior Software Engineer & Woman & 25-34 Years  & United States of America & 3-5 Years & Master's Degree\\
P9 & Data Governance Officer/Business Analyst & Woman & 25-34 Years & France & 3-5 Years & Doctorate Degree (Ph.D)\\
P10 & AI Engineer and Co-Founder & Man   & 25-34 Years & United States of America & 6-10 Years & Master's Degree\\
P11 & Principal Product Manager & Man  & 25-34 Years & France & 3-5 Years & Master's Degree\\
P12 & Full Stack Web Developer & Man  & 25-34 Years & Tunisia & 6-10 Years & Undergraduate Degree \\
P13 & Engineering Manager & Man & 25-34 Years & France & 6-10 Years & Master's Degree\\
P14 & Senior Research Software Engineer & Man & 25-34 Years & United States of America & 1-2 Years & Doctorate Degree (Ph.D)\\
P15 & Senior Software Developer & Man & 25-34 Years & Canada & 6-10 Years & Undergraduate Degree\\
\bottomrule
\end{tabular}
\end{table*}
In this section, we detail our methodology, including interview design, data collection, analysis, and theory building.

\subsection{Interview Design} 
\label{interviewDesign}
The purpose of the interview was to understand the motivations \textbf{(RQ1)}, the challenges \textbf{(RQ2)} faced by software engineers throughout their education and career journey and  the interplay of the two \textbf{(RQ3)}. To this end, we designed a semi-structured interview, to allow flexible exploration of participants' experiences while maintaining consistency across interviews. The interview script is available in the supplemental material \cite{suppdoc}. 



\subsection{Interview Pilot}
\label{interviewPilot}
The first author piloted the interviews with 3 participants, each of whom had experience in software engineering education and industry. After each pilot interview, the first two authors met to discuss and determine whether changes should be made. For instance, after the first pilot interview, a question was changed from ``What motivated you to pursue an education in software engineering?'' to ``What motivated you to build knowledge in software engineering?'', to account for participants who are self taught or did not pursue formal education in software engineering. Additionally, three questions surrounding career expectations were removed to keep the planned interview time in scope and under one hour. After two additional pilot interviews that did not prompt changes, we began recruiting participants for the interview study.

\subsection{Recruitment} 
\label{recruitment}
We recruited participants who had pursued a software engineering education and career. We used convenient sampling, recruiting participants within our networks via email. We contacted a total of 37 participants. Participants were sent a letter of information via email, which included a description of the study, data handling policies and researchers' contact information. Those interested in participating in the interview study were prompted in the email to schedule an interview via Microsoft Bookings. A total of 15 participants agreed to participate in the interview resulting in a 40\% response rate. Table \ref{table:participants} details the participants' information. Participants spanned a range of roles within software engineering including junior developers, senior engineers, architects, and technical leaders. The sample comprised 9 men and 6 women, primarily between 25-34 years of age. Participants were based across North America, Europe (France, Germany), and Africa (Tunisia). Their professional experience ranged from less than one year to over ten years, offering perspectives from early-career to senior professionals. Educational backgrounds were diverse, spanning technical training to doctoral degrees, with the majority holding undergraduate or master’s qualifications.

\subsection{Data Collection and Availability}
\label{dataCollection}
We arranged meeting dates and times according to the interviewee’s availability. All
interviews were conducted online via Microsoft teams. The first author conducted
the interviews. 
Before each interview, we obtained verbal consent from each participant for audio or video recording. 
Each interview lasted between 22 and 55 minutes (mean = 34 minutes) and was followed by a 3-minute demographic survey. 
After the 12th interview, no new themes or insights emerged. Past interview 12, we completed three additional interviews since they were already scheduled, and to provide further validation.The interview script, survey questions, and codebook can be found in the supplemental material \cite{suppdoc}. 

\subsection{Data Analysis}
We employed the Gioia Methodology \cite{gioia2013seeking} , an adaptation of grounded theory \cite{glaser1998grounded} 
to inductively and iteratively analyze the data. 
Despite its origins in organizational behaviour, the Gioia Methodology is increasingly adopted in software engineering research \cite{Russo2024Navigating} \cite{Knutas2022Civic} \cite{Verwijs2023ATheory}. For instance, \citet{Russo2024Navigating}, uses the Gioia methodology to derive a theoretical model of AI adoption in software
engineering. 

We qualitatively analyzed the transcripts by systematically reading the transcripts to develop a list of first-order concepts from excerpts using participant language without researcher interpretation (analogous to inductive open-coding in grounded theory \cite{glaser1998grounded}). We identified the motivations and challenges that each participant reported in both their education and career journeys. Next, we combined related concepts into more abstract second-order themes (analogous to  axial coding in grounded theory \cite{glaser1998grounded}). Finally, the researchers compiled the second-order themes into aggregate dimensions, used for theory development. The first and second authors met on a weekly basis to discuss
the grouped first-order concepts, refine definitions, and merge related second-order themes into representative aggregate dimensions. The first and second authors discussed the relationship between themes, and adjusted first- and second-order themes and until consensus. 



For instance, the second-order theme `Avoiding Consequences' became `Avoiding Alternatives' to better reflect the first-order concepts. Early themes of `Helping Others' and `Social Good' were combined to form `Impact'. Similarly, `Problem Solving' was moved from a first-order concept of `Curiosity' to its own second-order theme. The data analysis process took 26 weeks.

Following this analysis, we compiled the themes, concepts and dimensions into data structures that provide a transparent visualization of the transition from participant language to analytical dimensions. 
Finally, we used the data structures to develop a process model, using arrows to connect the various categories to show their temporal relationships. We iteratively improved the model by validating its accuracy in representing participant experiences. The authors discussed and iterated on the theory on a weekly basis. This process took 12 weeks. The model captures the dynamic interplay between motivations and challenges throughout the education-to-career continuum in software engineering, forming the foundation of our theoretical contribution. 

\section{Results}

\begin{figure*}
    \centering
    \includegraphics[width= \textwidth]{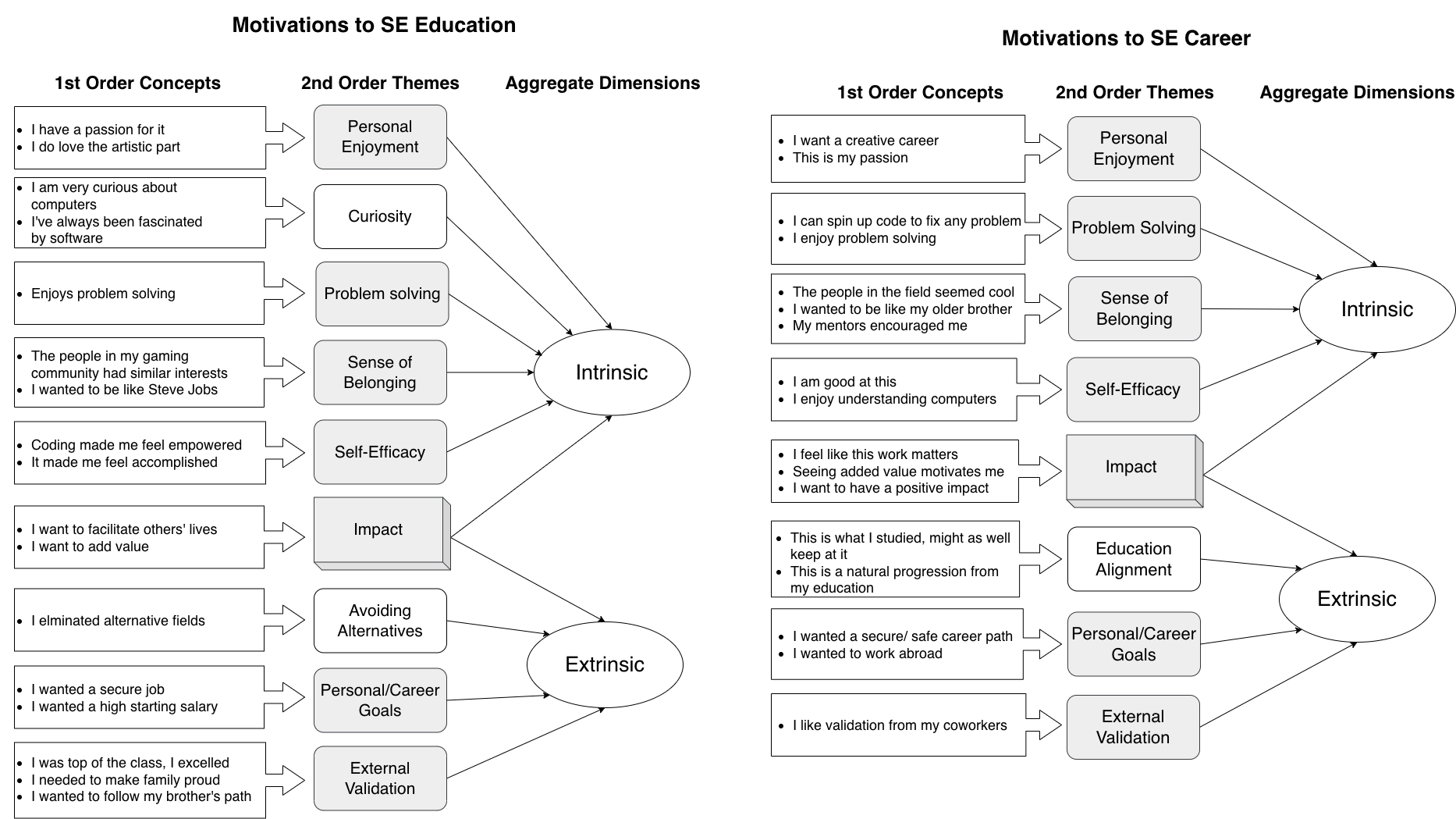}
    \caption{Data structures of motivations in software engineering education and career. The second-order themes and aggregate dimensions that are shared between education and career are highlighted in grey.}
    \Description{Two data structures illustrating how first-order concepts were grouped into second-order themes and aggregate dimensions for motivations across education and career.}
    \label{fig:Motivations}
\end{figure*}

In this section, we describe our findings on what motivates (see Section \ref{sec:Motivations}) and challenges (see Section \ref{sec:Challenges}) software engineers throughout their journey from education to practice and theorize the interplay of the two (see Section \ref{sec:theory}).


\subsection{Motivations in Software Engineering (RQ1)}
\label{sec:Motivations}
The data structure in Figure \ref{fig:Motivations} provides a visual representation of the analytical progression from first-order concepts to aggregate dimensions in both education and career motivations.

\subsubsection{Intrinsic Motivations}
This aggregate dimension captures the motivations to pursue software engineering that are rooted in an internal drive, rather than external rewards or consequences \cite{oudeyer2007intrinsic}. All intrinsic motivations except \textsc{curiosity} exist throughout the software engineering journey from education to professional practice, whereas \textsc{curiosity} relates exclusively to education.

\textbf{Personal Enjoyment:}
Participants frequently referred to an enjoyment for software engineering activities that drove their pursuit of the field. This motivation was found in both education (P1, P4, P6, P7, P11, P12, P13) and career (P1, P7, P10, P12, P13) aspirations. Participants who described personal enjoyment referred to three main motivations: Passion, Creativity and a Proximity to Other Interests.  One participant noted that their motivation was \inlinequote{purely passion driven, like it was fun (P4, Education).} Another participant noted their enjoyment as a natural driver of career choice, saying: \inlinequote{this is my passion. And as because it's my passion, then I should be doing that as a career (P1, Career).}

In addition to passion, participants also shared that they were motivated by the creative portions of software engineering. For instance, P7 described their enjoyment of their software engineering career,\inlinequote{which fuels the creativity (P7, Career).} Participants also related software engineering to other interests. As P11 shared, \inlinequote{It's got a nice puzzle aspect to it and I do enjoy like strategy and puzzle games (P11, Education).}

These experiences suggest that \textsc{personal enjoyment} plays a role in inspiring both educational and career decisions in software engineering and reinforces the broader dimension of intrinsic motivation.

\textbf{Curiosity:}
\textsc{Curiosity} emerged as a major educational motivator experienced by 14 out of 15 participants (P1, P2, P4, P5, P6, P7, P8, P9, P10, P11, P12, P13, P14, P15). Participants described an internally driven desire to understand how software systems work and to expand their knowledge beyond surface-level understanding. As one participant noted, \inlinequote{I would say most of all, curiosity. I was always just very curious about it. To me, it kind of seemed like, you know, magic (P11, Education).} Similarly, another explained, \inlinequote{somehow I was intrigued by making computers do things (P6, Education).} Another participant reflected this deeper epistemic curiosity: \inlinequote{I'm always curious at a certain level and not at a surface level. So when you show me something, I’ll always ask the what and the why underneath it (P7, Education).} Together, these perspectives highlight \textsc{curiosity} as an intrinsic motivation to software engineering education. 

\textbf{Problem Solving:}
\textsc{Problem solving} was another frequently expressed motivation to both education (P1, P2, P7, P8, P9, P10, P11) and careers (P7, P9, P10, P13). Some participants focused on the implementation aspect of \textsc{problem solving}. For example, one participant shared that they, \inlinequote{enjoy the challenge of trying to figure out a coding problem or figuring out how to implement something (P1, Education).} Another participant supported this view with a reflection on the internal nature of this motivator, \inlinequote{It [software engineering] was really a way to- to- create things to solve problems, to make things easier for myself then (P9, Education).}
A third participant described the satisfaction of problem solving through an analogy to artistic creation: \inlinequote{longing for the result also when I paint right. So, I am enjoying the painting process but this painting, like the longing for the final result that I can achieve (P11, Education).} In their career, P10 also describes this motivation, sharing, \inlinequote{ I want to solve the problems (P10, Career).} These accounts illustrate an intrinsic motivation for \textsc{problem solving}, where participants derived enjoyment from overcoming challenges and creating solutions. For some, the pleasure resided in the process of tackling complex problems, while for others, it was in the outcome, the sense of completion or mastery achieved when a solution worked. 



\textbf{Sense of Belonging:}
\textsc{Sense of Belonging} refers to the feeling of kinship to or within a perceived community \cite{allen2021belonging}. This motivation is observed in the pursuit of both education (P4, P8, P15) and career (P9, P10, P13). Motivations related to \textsc{sense of belonging} included seeking like-minded people, and following in the footsteps of role models.  One participant shared their experience in seeking community, noting that their motivation was, \inlinequote{surrounding also yourself with people who share the same goals (P8, Education).} 

P9 attributed some of their motivation to wanting to follow in the footsteps of family members, sharing\inlinequote{when I was in high school he [brother] started studying at software engineering and I had another uncle who's a software engineer (P9, Education).}
In contrast, while some participants were motivated by proximal connections (i.e., people in their immediate circle or group membership), others spoke of specific public role models. For example, one participant recalled that, for them, \inlinequote{seeing people like Steve Jobs would be like a source of inspiration at the time (P6, Education).}

Participants report interest in associating with a group, and their shared views demonstrate a desire to belong to a community as a motivator. 

\textbf{Self-Efficacy:}
\textsc{Self-Efficacy} represents the internal belief in one's capability to accomplish a particular task \cite{bandura1994self}. Participants often described feeling empowered, or confident in their abilities, as  motivators to both education (P1, P3, P4, P7, P11, P12, P14, P15) and careers (P10, P12).  For instance, a participant who described their first time coding as an initial motivator, describes the experience to be empowering, sharing \inlinequote{but it's just like felt empowering ...like you can make machines do something (P6, Education).}

In contrast, some participants attribute parts of their motivation to a software engineering career to a general desire for competence. For example, P3 shared, \inlinequote{I like to be good at things, I guess (P3, Career).} P12 shares this motivation, describing a desire to be good at things and sharing that the, \inlinequote{thing that I am  good at, is building software (P12, Education)}.


\subsubsection{Extrinsic Motivations}
This aggregate dimension captures the motivations of participants driven by external benefits or consequences \cite{oudeyer2007intrinsic}. Two out of the three extrinsic motivations appear throughout the software engineering journey from education to practice (see Figure \ref{fig:Motivations}). \textsc{Avoiding Alternatives} and \textsc{Education Alignment} are exclusive to eduction and career respectively. 

\textbf{Avoiding Alternatives:}
Some participants noted that a motivator to software engineering education (P8, P9) was the avoidance of alternative educational paths. For instance, P8 recalled \inlinequote{In high school, you had to choose ... between two things actually, like tourism or technical background. So I was like, oh, I- I- don't really want to do tourism like for studies. So, I went to software engineer (P8, Education).}

When presented with career options, participants selected software engineering as opposed to other, less-appealing alternatives. Interviewees reported this process-of-elimination method as a motivation for their pursuit of a software engineering education.

\textbf{Education Alignment:}
A number of participants attribute part their choice to pursue a career in software engineering to the fact that they had already pursued a related educational pathway (P3, P4, P6, P8, P9, P15). This theme refers to motivations stemming from the natural alignment between a participants' selection of a software engineering career and an education that they have already undertaken. One participant described their career pursuit as the result of an academic momentum because, \inlinequote{this is what I studied. Um, I might as well keep at it (P15, Career).} Another participant had a slightly different perspective, suggesting that pursuing a software engineering career was 
not a second choice 
but that it was simply a continuation of the same choice. P4 shared: \inlinequote{I don't think I ever did pursue it as a career. I think it just happened (P4, Career). }

These examples suggest that participants who pursued software engineering careers continued down the path due to a perceived natural progression.

\textbf{Personal/Career Goals:}
14 out of 15 participants noted \textsc{personal/ career goals} as a motivation in their education (P2, P4, P5, P6, P9, P13, P14, P15) or career (P1, P2, P5, P6, P9, P10, P11, P13, P14). Participants shared varied personal/career goal motivations, including financial gain, job security and international mobility. One participant states that, \inlinequote{[software engineering is] also a field which is paying good, in which there's also a lot of opportunities (P13, Education).}, highlighting the financial benefits that motivated them. P10 corroborates financial benefit as a career motivation where \inlinequote{obviously there is like a lot of people are earning too (P10, Career).}

Additionally, 11 participants noted that their motivations changed over time to encompass personal/ career goals. For example, P14 shared later in their journey that, \inlinequote{As I got older, it was more so about being able to provide stability and security for myself in my life, being able to get that, you know, that job that can able that can provide those things (P14, Career).} Additionally, P11 reflected on their change in career motivations, \inlinequote{I was potentially more motivated or interested in the concept of entrepreneurship rather than software engineering directly, and software engineering was potentially an enabler for me in that entrepreneurship (P11, Career).}, showing a focus on the prospective career benefits over software engineering itself.

\textbf{External Validation:}
Certain participants reported pursuing \textsc{external validation} as a motivation to both education and careers. Interviewees noted validation from family members, academic institutions and co-workers. Though similar to \textsc{self-efficacy}, which is an internal confidence/ appreciation of ones' skills, the \textsc{external validation} motivation (also known as others' approval \cite{crocker2003contingencies}) presents a distinct difference in that individuals seek an exterior source of recognition. As one participant states, \inlinequote{I love studying to get good grades. cause I love the academic validation (P4, Education).} Another participant shared a similar view, \inlinequote{
I felt like I needed to go into stem [...] because my most of my family was in stem and it felt just the standard to make them proud (P3, Education).} The same participant shared, \inlinequote{It just goes the same for work as well, like validation from your coworkers and your manager (P3, Career).}, indicating a continuation of a desire for recognition beyond an educational context. These experiences suggest the existence of \textsc{external validation} as continuous extrinsic motivation that span the careers of certain software engineers.

\subsubsection{Impact:}
\textsc{Impact} was noted several times as a motivation by participants across both education (P11, P13) and especially in their careers (P4, P10-P14). Interviewees often shared an internal drive to do meaningful work, describing a desire to \inlinequote{feel like, you know, though the work that I am going to do is going to matter (P14, Career).} In educational contexts, participants shared similar motivations. For example, P13 recounts that \inlinequote{it all starts from, 
... facilitating people's life in the end. That's I think a very important aspect of why I choose software engineering (P13, Education).} This highlights that the desire to have an impact is originating internally. However, they also point to extrinsic motivation due to the influence of external outcomes. For example, P11 described an extracurricular project that allowed them to make an impact saying, \inlinequote{ I was thinking of the customer experience, of ... the added value to the customer (P11, Education).} Given the mixed nature of the impact-related discussions within the data, we consider \textsc{Impact} to relate to both intrinsic and extrinsic aggregate dimensions (see Figure \ref{fig:Motivations}).  

\begin{figure*}
    \centering
    \includegraphics[width= \textwidth]{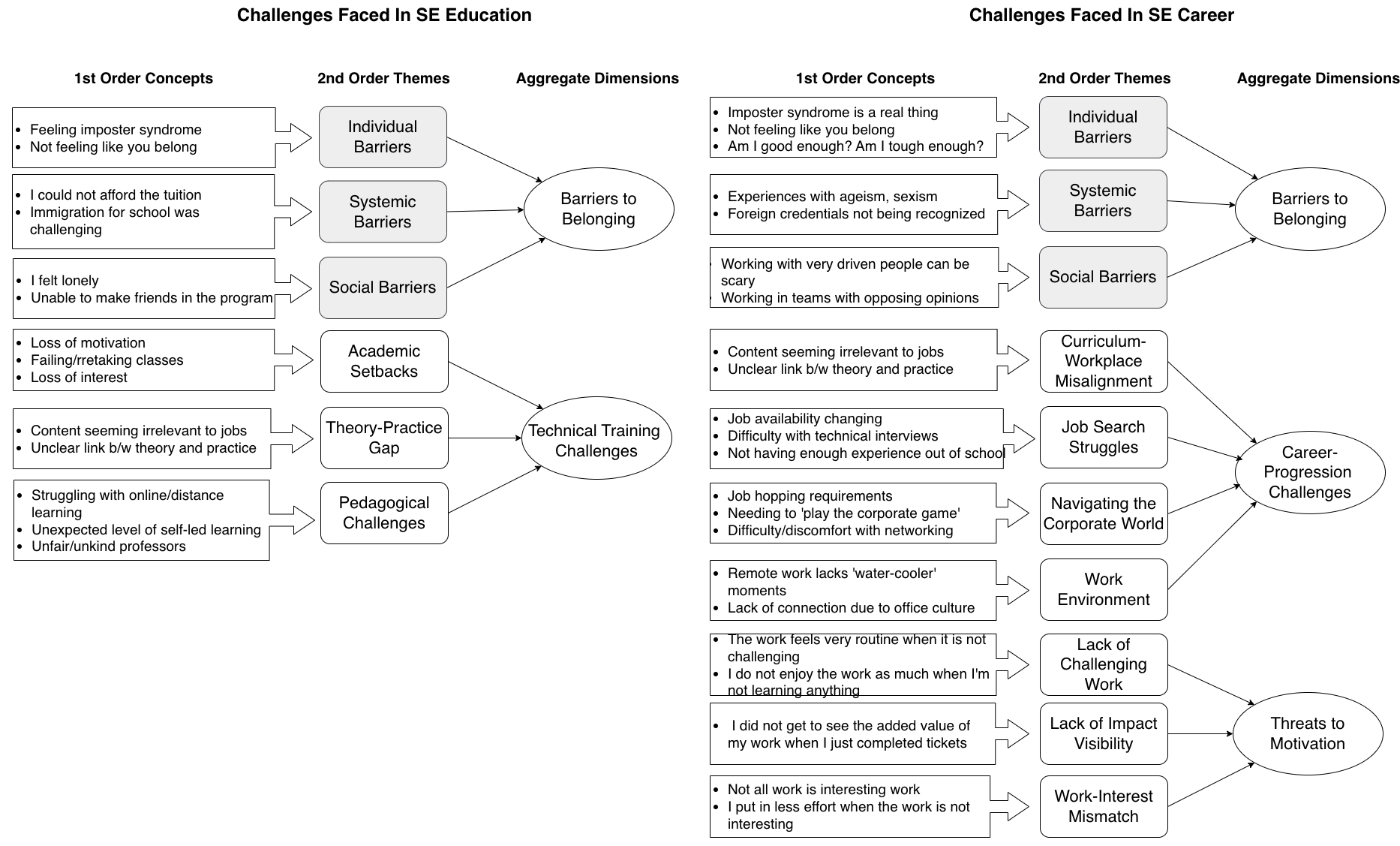}
    \caption{Data structures of challenges faced in software engineering education and career. The second-order themes and aggregate dimensions that are shared between education and career are highlighted in grey.}
    \Description{Two data structures illustrating how first-order concepts were grouped into second-order themes and aggregate dimensions for challenges across education and career.}
    \label{fig:Challenges}
\end{figure*}

\subsection{Challenges in Software Engineering (RQ2)}
\label{sec:Challenges}

Our results indicate that \textsc{barriers to belonging} can arise during both the educational and professional stages of a software engineering journey (see Figure ~\ref{fig:Challenges}). The \textsc{technical training challenges} observed in education have a corresponding form in practice, reflected as \textsc{career progression challenges}. Notably, professional experiences also reveal threats to motivation including \textsc{lack of challenging work} and \textsc{Work-Interest Mismatch}.

\subsubsection{Barriers to Belonging:} \label{sec:barriers to belonging}
The only aggregate dimension that continues from education to career is \textsc{Barriers to Belonging}, which refers to individual, social, and systemic challenges faced that impact feelings of belonging within software engineering. 

\textbf{Individual Barriers:}
\textsc{Individual Barriers} are barriers to belonging which are grounded in personal beliefs or experiences. For example, imposter syndrome was commonly reported by participants (P3, P7), describing \inlinequote{Just imposter syndrome. Of course, being like, a woman in this field, you're like, am I smart enough in the way that the men are smart and my capable enough (P7, Education).} This sentiment is echoed by P3 in their early career, \inlinequote{it's the imposter syndrome thing, thinking that. Am I? I'm new to this field. Am I smart enough for it? Am I capable enough for it? Am I tough enough? Is a huge one (P3, Career).} 

In addition to imposter syndrome, P4 commented on their educational experience with learning disabilities, saying \inlinequote{I had learning disabilities and you know, I- ADHD and things that made school difficult for me (P4, Education).}

Both learning disabilities and imposter syndrome represent \textsc{individual barriers} faced by participants. Further, both participants who encountered imposter syndrome in education continued to experience these barriers later in their careers  (P3, P7). 

\textbf{Systemic Barriers:}
These challenges arise as a direct result of existing societal systems and norms. This describes barriers such as discrimination, lack of access to opportunities, and the under-representation of gender minorities. P3 echoed this last barrier reflecting on their educational experience, saying \inlinequote{I'd say it's definitely easier to make friends with people that you think you can relate to, and there's not a lot of woman in computer science, as you know, very few girls in the classes (P3, Education).}

We also identified \textsc{systemic barriers} that persist throughout participants' careers. Two participants (P8, P11) described difficulties related to international credential recognition, where degrees or grades from reputable institutions in one country were not acknowledged elsewhere. As one participant explained,
\inlinequote{In my home country, we are in one of, like, say, one or two percent potentially best engineering schools. But when we're outside in the world, this is an unknown school at the end (P11, Career).} Another participant shared their frustration when applying to graduate programs that used a different grading scale: \inlinequote{I calculated my grade based on the qualitative [grade] from my university that was approved. The department accepted me, but then the university rejected me 
(P8, Education)}.

These experiences illustrate how institutionalized evaluation systems can inadvertently disadvantage individuals whose educational backgrounds originate from different academic or geographic contexts, thereby perpetuating inequities in global mobility and career advancement.

\textbf{Social Barriers:}
\textsc{Social Barriers} describe challenges that stem from interpersonal dynamics and social environments. Some participants reflected on the isolation that they experienced in their education (P2, P9) and career (P1, P4, P6, P12). For instance, one participant shared that \inlinequote{Making connections with my peers and finding friends in classes. That one was hard and sometimes just feel alone in class and it's harder to like stay motivated (P2, Education).} Another participant emphasizes this challenge in a career context, sharing having \inlinequote{a colleague that I pretty much challenged my ideas with or you know, but it was pretty much a lonesome career (P9, Career).}

Further in careers, a number of participants reported experiencing difficulties in working with others (P1, P4, P6, P9, P12). For instance, P12 highlights their experience in working with different types of personalities, where \inlinequote{There are some that are open and accepting... Some ... even if the whole team agrees on doing things in a certain way, they they resist that (P12, Career).}

\subsubsection{Technical Training Challenges: }
\label{Technical Training}
Our result shows that \textsc{technical training challenges} occurred frequently in software engineering education (P1-P3, P5-P10, P12-P15). This refers to obstacles related to all aspects of technical training, including \textsc{academic setbacks} (P2, P5, P6, P15), the \textsc{theory-practice gap} (P3, P5, P7, P9, P12, P13) and \textsc{pedagogical challenges} (P1-P3, P5, P8, P10, P14). 

\textbf{Academic Setbacks:}
This theme indicates challenges related to setbacks within a participant's education which delay or negatively impact their academic journey. Participants who have experienced academic setbacks mentioned burnout (P2, P6) and failed classes (P2, P5) as common challenges. For example, P2 shared their experience with failed classes noting that, \inlinequote{there's always the struggles of like the hard classes that you have to retake (P2, Education).} Another participant reflected on the financial and temporal impact of failed classes, as they shared \inlinequote{I had the family support and the financial support to be able to do five years instead of four in my education. Like, if I didn't have those things, I might not have been able to do it 
(P5, Education).}

\textbf{Theory-Practice Gap:}
The \textsc{Theory-Practice Gap} refers to the experience of students who do not see a clear alignment between the content that they are learning and their intended applications. Several participants shared their experience with this educational challenge (P3, P5, P7, P9, P12, P13). For instance, one participant shares that, \inlinequote{The only challenge maybe that I found was the two year, first year 
I didn't get why we would have to study math and physics (P12, Education).} Similarly, a second participant emphasized the lack of clarity between their learning and their intended implementation, sharing, \inlinequote{I think that's one of the biggest challenge that I encountered while I was studying is just to like bridge the gap between. The theory and actually the implementation part of it (P7, Education).}

\textbf{Pedagogical Challenges:}
Several participants (P1-P3, P5, P8, P10, P14) reported challenges related to their educational experience (i.e., teachers and learning environments). The need for self-led learning was frequently noted as a challenge among participants where \inlinequote{you had to kind of figure it out on your own (P1, Education).} Some have identified it as \inlinequote{one of the biggest challenges that you don't necessarily always feel like there's enough support ... like go out of my own way to learn those things 
(P14, Education).}

In addition to self-led learning, participants also noted challenges relating to teachers. For example, P2 describes non motivating communication from teachers, sharing \inlinequote{there was like a lot of like what I felt was like passive aggressive messages (P2, Education).} 

Participants mentioned another challenging component of their educational experience: weed out classes. P5 described these classes as being purposefully difficult as to deter students from continuing in the program: \inlinequote{one of the big problems is the fact that it very much felt like the program was trying to weed out a lot of students (P5, Education).}



\subsubsection{Career-Progression Challenges:}
\label{CareerProgression}
\textsc{Career-Progression Challenges} (See Figure \ref{fig:Challenges}) were the most frequently experienced career related challenges mentioned by 14 out of 15 participants (P1-P11, P13-P15). Career-Progression Challenges are obstacles that limit intended career growth within software engineering namely \textsc{Curriculum-Workplace Misalignment}, \textsc{Job Search Struggles}, \textsc{Navigating the Corporate World}, and challenges related to the \textsc{Work Environment}. 

\textbf{Curriculum-Workplace Misalignment: } 
This challenge parallels the educational \textsc{theory–practice gap'} but manifests in professional settings, reflecting a misalignment between academic preparation and the practical demands of software engineering work. Participants shared that the roles they are applying to often have unanticipated expectations, as P9 states, \inlinequote{When people hire, they expect you to do 'ABC' and you maybe know 'A' have an idea about 'B', but do not know 'C' at all. And you have to cope with what you're expected to do (P9, Career).}, highlighting the gap between their knowledge and the expectations for prospective jobs. 

P4 shared a different view, focusing more on their unpreparedness for the non technical skills or \inlinequote{the politics of things (P4, Career).} This participant went on to share that they \inlinequote{ I am not prepared for and still find myself unprepared for (P4, Career).} These results highlight a disconnect between the workplace requirements that software engineers are experiencing, and their educations.

\textbf{Job Search Struggles: }
This challenge describes negative experiences related to finding software engineering related employment. Our findings show that 8 out of 9 participants who experienced \textsc{Job Search Struggles} in their career, also experienced \textsc{Technical Training challenges} in their education. 
Participants point out the lack of industry experience as a main hurdle in their first job search, sharing that \inlinequote{Fresh out of out of university, they all want you to have job experience (P15, Career)}.
Participants also shared challenges related to methods of assessment in job interviews that they deemed \inlinequote{not as useful for evaluating my skills and ability (P1, Career).} Indeed, P1 further explains that \inlinequote{the ones [challenge] that immediately comes to mind is coding problems like Leet Code ... it's this arbitrary coding problem is really, really my kryptonite (P1, Career). }. Similarly, participants noted that   staying up-to date on new technologies  posed a challenge during the job search process. For example, P10 describes a \inlinequote{deep initial learning curve (P10, Career)} that exists when learning new frameworks. They continue, sharing that early in their career \inlinequote{hesitated to even get into [new frameworks]}.

Moreover, participants found that they did not adequately understand the pathway to a job after their education. For instance, P8 shared that  \inlinequote{one of the challenges is not having like a very clear path afterwards}. 


\textbf{Navigating the Corporate World: }
Unlike \textsc{job search struggles}, this theme captures challenges related to \textsc{navigating the corporate world}, where participants described difficulties adapting to organizational expectations and workplace dynamics. Participants refer to difficulties with the non-technical requirements within their roles (P1, P3-P6, P9, P10). These difficulties act as moderators to an engineer's ability to progress in their role. For instance, P4 shared hurdles related to the social expectations within a corporate environment: \inlinequote{I get like a performance review: Oh well, not enough people know you ... And it's taken me a long time to understand what that meant or not enough people can talk about working with you.
So I think that that has been something I've I've had to learn to do. And as an introvert, it's not my natural (P4, Career).}

Another participant mentioned a challenge associated with securing a salary increase. Specifically, P5 describes that in their experience, \inlinequote{if you want salary increases, you gotta job hop (P5, Career).}

\textbf{Work Environment:}
This theme refers to challenges involving working hours (P2) and remote work challenges (P2, P4, P5). One participant shared their experience with remote work, recounting \inlinequote{It's [working remotely] harder as well, right? ... You don't like, have those water cooler moments 
(P4, Career).}
Similarly, P2 reflected on their role's availability requirements where  \inlinequote{The tech lifestyle is less so, like the nine to five office and more like being available all the time (P2, Career)}. 

Naturally, this challenge occurs during career pursuit, and manifests as a constraining influence on Career/ Personal Goals (see Figure \ref{fig:Model}).

\subsubsection{Threats to Motivation: }
\label{ThreatsMotivation}
While we find all challenges observed as having a constraining influence on motivation, this aggregate dimension represents challenges that manifest with a direct, negative impact on a specific second order motivation (See section \ref{sec:theory}). Specifically, the \textsc{Lack of Challenging Work} challenge is directly related to the \textsc{Problem Solving} motivation, the \textsc{Lack of Impact Visibility} challenge is associated with the \textsc{Impact} motivation, and \textsc{Work-Interest Mismatch} impacts the \textsc{Personal Enjoyment} motivation. We only observed the occurrence of these challenges at the career-level, and find them to have an effect on career-level motivations.
\begin{figure*} 
\centering
    \includegraphics[width= \textwidth]{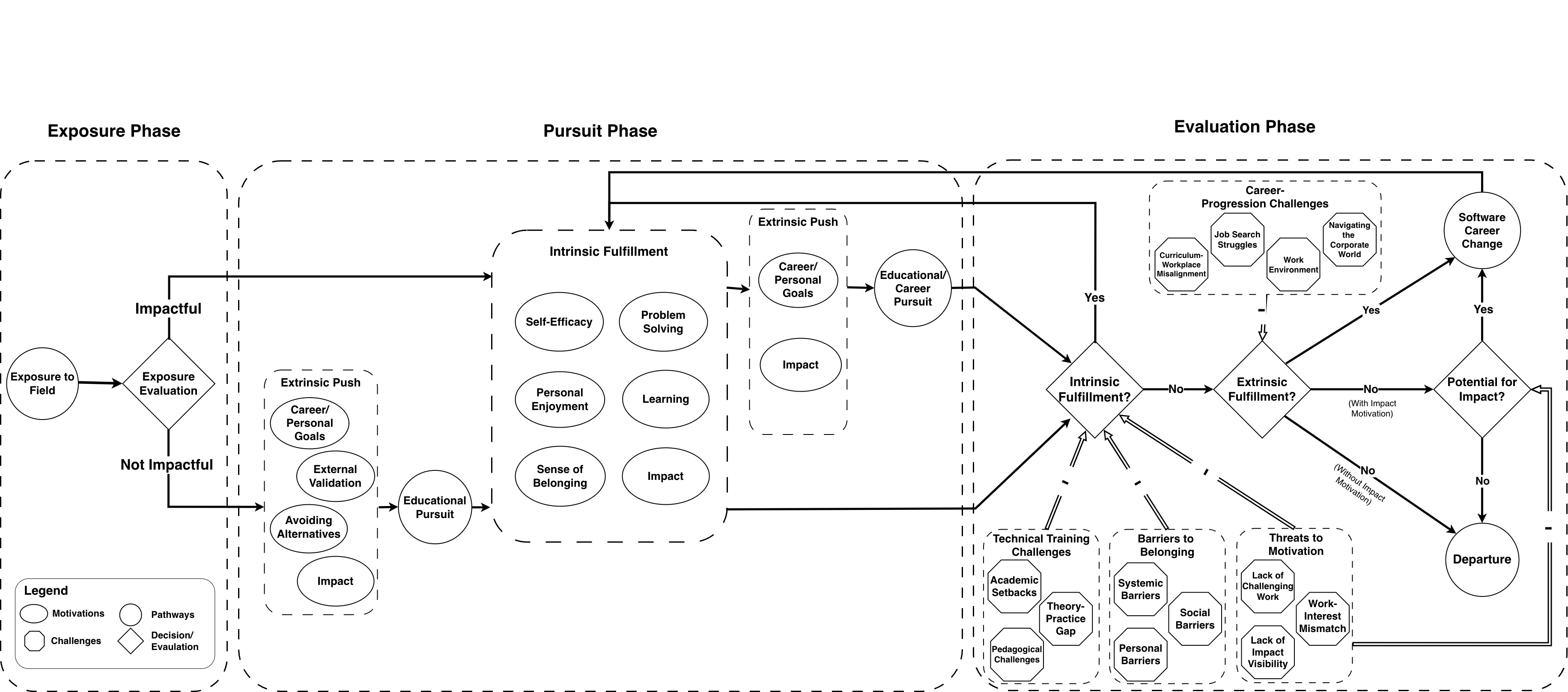}
    \caption{Exposure-Pursuit-Evaluation (EPE) process model of motivations and challenges in software engineering.}
    \Description{Three-phase process model of motivation and challenges in software engineering, showing how exposure, motivation, and challenges interact to shape software engineering careers.}
    \label{fig:Model}
\end{figure*}

\textbf{Lack of Challenging Work: }
Several participants reported a lack of challenging or engaging work as a barrier faced within their career(P11, P15). P11 described their experience with a lack of challenging work, sharing \inlinequote{maybe I'm just not having the technical challenge that I want (P11, Career).} Indeed, they continue by sharing their reasoning for a career change: \inlinequote{maybe I need to find something more challenging technically (P11, Career)}.
This highlights that a lack of challenging work can reduces a software engineer's intrinsic motivators hindering interest and enjoyment within the field.

\textbf{Lack of Impact Visibility: }
This challenge refers to a software engineer's inability to view the positive impacts of their work, and the inability to produce their intended impact.
We find this challenge to be directly related to the \textsc{Impact} motivation, and to only emerge for participants who posses the \textsc{Impact} motivation. One participant who spoke strongly of Impact as a motivator to both education and career, reflected on their experience as a full-stack developer \inlinequote{It wasn't necessary fulfilling for me, right? Because in the sense that it was, you are doing kind of specific tickets, specific developments you ask for, you don't really understand why or what the added value for them 
(P11, Career)}. 
As illustrated in the quote, we find that an inability to perceive one’s impact can negatively affects career motivation.

\textbf{Work-Interest Mismatch: }
This theme refers to a misalignment between a software engineer's interests and their work. \inlinequote{Challenges is maybe also making that work for me for my interests. Obviously, if you were in a company, it's not necessarily always the work you want to do (P6, Career).} Another participant describes their experience, sharing that without interesting work, \inlinequote{I get too comfortable ... I don't feel like I'm learning anything new ... I put in less effort (P15, Career).} P15's experience highlights that \textsc{work-interest mismatch} can impact \textsc{personal enjoyment} motivation (see Figure \ref{fig:Model}).

\subsection{Exposure–Pursuit–Evaluation (EPE) Model}
\label{sec:theory}
\subsubsection{The Theory:}
Following data analysis, we utilized each data structure (Figures \ref{fig:Motivations} and \ref{fig:Challenges}) to develop a process model of the relationships between motivations and challenges in software engineering. We then used the model, as seen in Figure \ref{fig:Model}, to develop our theory, which we summarize as follows: 

\MyBox{ \textit{Intrinsic motivation in software engineering develops through impactful exposure to the field, while individuals for whom exposure was not impactful will rely on an extrinsic push to pursue the field. Both types of motivation can be fulfilled through career and educational advancement, or limited by challenges encountered during their pursuit. When challenges prevent motivation fulfillment, individuals may shift roles within software engineering or leave the field entirely.}}


\subsubsection{Exposure–Pursuit–Evaluation (EPE) Model}
We propose the Exposure–Pursuit–Evaluation (EPE) Process Model to explain how motivations and challenges interact across the software engineering education–career continuum. The model is divided into three phases that each reflect motivations and challenges across education and career: (1) Exposure Phase, (2) Pursuit Phase and (3) Evaluation Phase. Preceding sections outlined the components of the model, the following sections describe the connections between them.

\textbf{Exposure Phase: }
This first phase marks an individual's first exposure to software engineering. For example, four participants (P4, P7, P11, P14) reported receiving a gifted computer as their first exposure, while two participants (P6, P12) describe their first exposure as a transition from a related interest (gaming, animation, etc.). Depending on whether their initial exposure is perceived as impactful, individuals take one of two pathways, which are explored in the pursuit phase (see Figure \ref{fig:Model}). We define impactful exposure as an experience through which a participant developed an internal drive to software engineering. An example of an impactful exposure to the field is that of P7 that \inlinequote{just like felt empowering (P7).}

\textbf{Pursuit Phase:}
This second phase describes an individual's pursuit of the software engineering field. First, if the exposure phase resulted in an impactful exposure, we find that individuals directly enter a state of intrinsic fulfillment. We define intrinsic fulfillment as the feeling of having one or more intrinsic motivators satisfied. For example, feeling capable (\textsc{self-efficacy}) or enjoying the related activities (\textsc{personal enjoyment}). Following this intrinsic motivational fulfillment, an individual will go on to pursue a software engineering education through an extrinsic push. The term extrinsic push refers to an externally-based motivations to pursue the field. 

We find that these individuals experience intrinsic motivation as a primary motivator, and extrinsic motivations as a secondary motivation. 

In contrast, we find that those who do not have an impactful exposure have the inverse experience of extrinsic motivation as a primary motivator and build their intrinsic motivation throughout their educational pursuit. For example, P3 describes their non-impactful exposure, and their extrinsic push saying, \inlinequote{Actually, I didn't enjoy high school computer science. It just felt like the easiest option of all the STEM fields, and I felt like I needed to go into STEM ... it felt just the standard to make [my family] proud (P3, Education).} While individuals without impactful exposure may require the extrinsic push, we find that they end up developing intrinsic fulfillment during their educational pursuit.

\textbf{Evaluation Phase:}
Following the pursuit phase, we find that individuals enter an evaluation phase. In this  phase, individuals experiencing intrinsic fulfillment will continue in their pursuit. Those not experiencing intrinsic fulfillment will evaluate their extrinsic fulfillment to determine if a change of career to a related area may be required. Finally, if extrinsic fulfillment is also missing, individuals without an impact motivation depart, and individuals for whom impact is a motivator will evaluate the potential for impact before deciding whether to make a software career change or depart the field.

Moreover, we propose that challenges moderate the evaluation stage. Technical Training Challenges and Barriers to Belonging limit intrinsic fulfillment. Career-Progression Challenges limit extrinsic fulfillment, and Threats to Motivation limit both intrinsic fulfillment and perceived impact potential.

\section{Discussion}
\label{sec:discussion}
In this section, we assess our results, and discuss implications for educators and companies.

\textbf{Initial Exposure} plays a major role in decisions to pursue and persist within software engineering. We found a variety of initial exposures to the field. Several exposures were shared among participants; however, not all participants found them to be equally impactful. For example, P2 and P3 both described exposures within high school computer science courses, but unlike P2, P3 did not find the experience to be impactful. Despite this, participants without impactful exposure were still able to find intrinsic motivation development. Providing a wide array of exposures to software engineering could support more engineers to develop stronger intrinsic fulfillment.

\textbf{Motivation Development} follows initial exposure. Software engineering's external incentives, such as flexible work environments and competitive compensation are capable of attracting people to the field. However, extrinsic benefits alone are not enough to encourage persistence. Self-Determination Theory \cite{deci1980self} states that intrinsic motivations are stronger, and persists even when extrinsic rewards are removed. Thus, cultivating intrinsic motivation is crucial for encouraging long-term persistence within a field.

Those who begin without developing an intrinsic appreciation for software engineering often develop one during their educational pursuit. Specifically, fulfillment of the sense of self-efficacy could be supported with more diverse teaching and evaluation methods. Bandura et al. found that self-efficacy influences performance and persistence outcomes \cite{bandura1994self}. Additionally, self-efficacy is not fixed, so early wins and successes can aid in building persistence. 

Similarly, persistence of individuals could be influenced by improving personal enjoyment, particularly in the connection of software engineering to other interests. The field of software engineering has a broad array of applications, allowing individuals to connect their diverse interests to the discipline. 

\textbf{Evaluation and Persistence} were found to be moderated by challenges, with those experiencing challenges that overshadowed their motivational fulfillment being more likely to consider changing careers or departing from the field entirely. The evaluation challenges shared by engineers were often ongoing difficulties for which they did not posses the tools to overcome.

Further research could seek to understand what constitutes the successful mitigation of these challenges or develop tools to aid software engineers in overcoming them. 

\textbf{Broader Implications:}
Both educators and companies can play a crucial role in how individuals form and maintain their motivation in software engineering. For educators, designing impactful exposure to the field of software engineering is critical. Given the individual nature of evaluating exposure, institutions should aim to diversify the way that students encounter software engineering, with a special focus on highlighting real-world impact. These experiences can allow individuals to build motivation internally, rather than relying solely on extrinsic motivations for their initial pursuit. Moreover, educational programs can also encourage self-efficacy in students by diversifying evaluations and teaching, enabling students more opportunities for early 'wins'. 

For companies, retaining software engineers should go beyond external rewards. While these incentives, such as competitive pay and flexible work options are shown to draw people to the field, ultimately persistence depends on intrinsic motivation and potential for impact. Fostering work environments that support intrinsic fulfillment could look like supporting continuous learning, aligning tasks with personal interests or values, and allowing engineers to see the outcomes of their work. Further, companies can develop support systems such as mentorship programs to help individuals navigate the challenges that they may face.

\section{Threats to validity}
\textbf{Construct validity} in qualitative research concerns the accurate definition and representation of constructs. One potential threat arises from asking poorly formulated questions. To address this, 
we piloted the interview script with three participants external to the research team. Another concern involves the coding process, which may introduce bias or misinterpretation. To mitigate this, we compared emerging first-order themes with the existing ones \cite{barney2017discovery} and held regular meetings with the research team to discuss and clarify themes. The final set of codes was reviewed and agreed upon by the entire team. To strengthen credibility, we present our coding data structures in Figures \ref{fig:Motivations} and \ref{fig:Challenges}

\textbf{Internal validity} relates to how accurately the study captures participants’ realities, in our case, their motivations and challenges throughout their software engineering journey from education to professional practice. Our sample comprised 15 participants with diverse roles, genders, and educational backgrounds (see Table \ref{table:participants}). 
This diversity enhanced our ability to capture varied experiences within software engineering practice. After the 12th interview no new themes or insights emerged. 

\textbf{Reliability} concerns the degree to which our findings can be reproduced. In general, qualitative research is challenging to replicate due to the evolving nature of human behaviours, and perceptions. However, we ensured consistency by continually comparing our analysis with existing first-order themes and the first two authors conducted weekly meetings to review and refine first- and second- order themes until reaching consensus.

\textbf{Theoretical saturation.} A potential threat to validity concerns whether theoretical saturation was achieved. In qualitative research, the quality of participants’ insights, rather than the number of participants, strengthens confidence in the findings. We continued conducting interviews until no new codes emerged (12 interviews) and then carried out three additional interviews, which yielded no new constructs. While we do not claim to have reached full theoretical saturation, the collected data provided a coherent and comprehensive understanding of software engineers’ motivations and challenges across their careers and professional practices, as well as how these aspects interact.

\section{Background and Related Work}
Prior research has examined software engineers’ motivations largely from an organizational perspective, linking them to company outcomes. A seminal review by \citet{beecham2008motivation} synthesized early studies on motivation in software engineering and observed that SE motivation is context-dependent
and noted that existing models were disparate and didn’t account for changing roles/environments. Building on this foundation, \citet{sharp2009models} developed a motivation model grounded in job characteristics and satisfaction. Similarly, \citet{Cesar2012Towards} modelled motivation and job satisfaction within a Brazilian government organization, and \citet{FrancaMotivation2020} extended this line of work through a multi-case study across four Brazilian companies to theorize work motivations in SE contexts. Complementing these, \citet{verner2014factors} identified culturally independent factors influencing motivation in software engineering teams.

Motivation has also been explored in specific subpopulations, such as open-source contributors \cite{gerosa2021shifting, alexander2002working}, adult women in software engineering \cite{hyrynsalmi2019motivates}, and women in STEM more broadly \cite{Bustamante2021Motivation}. For example, \citet{hyrynsalmi2019motivates} found that adult women's transitions to software careers were mitigated by obstacles including a lack of career mentorship and unclear educational pathways. Other strands of research have investigated how motivation shapes practice adoption \cite{song2018promotion} and project success in agile teams \cite{salman2021empirical}, as well as how it is affected by stress \cite{suarez2024stress} and global team collaboration \cite{noll2017motivation}.

From a challenges perspective, prior work has examined specific difficulties encountered in software engineering careers, including burnout \cite{tulili2023burnout, trinkenreich2023model, paula2024burnout}, imposter phenomenon \cite{chen2024impostor, rosenstein2020identifying, guenes2024impostor}, unhappiness \cite{graziotin2017unhappiness, graziotin2017consequences, obi2025identifying}, and the industry–skill gap \cite{diniz2024skill}. For instance, \citet{guenes2024impostor} found that nearly half of their participants reported frequent imposter symdrome experiences which disproportionately affecting women and racialized individuals.

Challenges have also been studied within specific subpopulations, including blockchain developers \cite{bosu2019understanding}, open-source contributors \cite{guizani2021long, steinmacher2015social}, women in software engineering \cite{oliveira2025investigating, canedo2019barriers, guzman2024mind, boman2024breaking}, and engineers on the autism spectrum (ASD) \cite{costello2021professional, Kiev2025Cognitive}. For example, \citet{costello2021professional} reported that social aspects of software engineering work—such as stand-up meetings and nuanced communication—can present particular challenges for some individuals with ASD. In a systematic literature review, \citet{canedo2019barriers} find that the under-representation of women in software engineering may be related to challenges with the work environment, a result of male gender-bias.

These studies provide insights into specific facets of motivation and challenges in software engineering. However, they have largely examined one of the two dimensions in isolation. Prior research has focused either on what drives software engineers or on what hinders them, and often within limited contexts such as education or the workplace alone. Consequently, little is known about how motivations and challenges co-occur, interact, and evolve across the broader journey from learning to professional practice. This study addresses that gap by developing an integrated, process-based theory explaining how these forces jointly shape persistence and adaptation across the software engineering education–career continuum.

\section{Conclusion}
\label{sec:conclusion}
In this work, we theorized how motivations and challenges jointly shape software engineers’ pathways from education to practice. Drawing on 15 semi-structured interviews and an inductive analysis adapted from organizational behaviour, we derived the Exposure–Pursuit–Evaluation (EPE) process model and taxonomies of motivations and challenges. Our findings show that intrinsic and extrinsic motives operate together, while barriers including belonging, training, and career progression constrain fulfillment and can precipitate role changes or departures. Because intrinsic drivers are more durable than extrinsic incentives, educators and organizations should prioritize conditions that build intrinsic fulfillment (e.g., early wins that strengthen self-efficacy, meaningful problem solving, visible impact) and aim to remove limiting barriers (e.g., misalignment between curriculum and workplace, inequities that undermine belonging). Practically, this implies revisiting assessment and pedagogy, making impact visible at work, and supporting navigation of corporate expectations. The EPE model offers a lens for diagnosing where interventions will matter most along the education–career continuum and provides a grounded basis for future longitudinal and interventional studies.

\bibliographystyle{ACM-Reference-Format}
\bibliography{biblio}
\end{document}